\documentclass[12pt]{article}
\usepackage{jhep-mod}
\usepackage{bm}
\usepackage{amssymb, amsmath, amsthm, physics}
\usepackage{mathrsfs}
\usepackage{hyperref}
\usepackage{multicol}
\usepackage{float, array, tabu}
\usepackage[tableposition=top]{caption}
\usepackage{tocloft}% http://ctan.org/pkg/tocloft
\usepackage[utf8]{inputenc}
\usepackage{enumerate}
\usepackage{bigints}
\usepackage{appendix}
\usepackage{graphicx}
\usepackage{float}
\usepackage{tikz}
\usepackage{setspace}
\usepackage{cancel}
\usepackage[normalem]{ulem}
\definecolor{purple}{rgb}{1,0,1}
\definecolor{lime}{HTML}{A6CE39} % needs xcolor
\renewcommand{\r}{\hat{r}}
\renewcommand{\t}{\hat{t}}
\renewcommand{\th}{\hat{\theta}}
\newcommand{\ph}{\hat{\phi}}
\definecolor{lime}{HTML}{A6CE39}
\newcommand{\orcidicon}{%
	\begin{tikzpicture}
	\draw[lime, fill=lime] (0,0) 
		circle [radius=0.16] 
		node[white] {{\fontfamily{qag}\selectfont \tiny ID}};
	\draw[white, fill=white] (-0.0625,0.095) 
		circle [radius=0.007];
	\end{tikzpicture}
	\hspace{-5mm}
}
\newcommand\orcidThomas{{\href{https://orcid.org/0000-0002-0314-4136}{\orcidicon}}}
\newcommand\orcidAlex{{\href{https://orcid.org/0000-0002-1763-3563}{\orcidicon}}}
\newcommand\orcidMatt{{\href{https://orcid.org/0000-0003-1088-6485}{\orcidicon}}}
\begin{document}
%========================================================
%========================================================
\title{\vspace{-25pt}
\huge{
General class of ``quantum deformed'' regular black holes\\
}
}
%========================================================
%========================================================
\author{
\Large
Thomas Berry\orcidThomas\!,
Alex Simpson\orcidAlex\!, {\sf{and}}
Matt Visser\orcidMatt}
%========================================================
%========================================================
%========================================================
%========================================================
\affiliation{School of Mathematics and Statistics, Victoria University of Wellington, \\
\null\qquad PO Box 600, Wellington 6140, New Zealand.}
%========================================================
%========================================================
\emailAdd{thomas.berry@sms.vuw.ac.nz}
\emailAdd{alex.simpson@sms.vuw.ac.nz}
\emailAdd{matt.visser@sms.vuw.ac.nz}
%========================================================
%========================================================
\parindent0pt
\parskip7pt

\abstract{

\noindent
We discuss the ``quantum deformed Schwarzschild spacetime'' as originally introduced by Kazakov and Solodukhin in 1993, and investigate the precise sense in which it does and does not satisfy the \emph{desiderata} for being a ``regular black hole''. We shall carefully distinguish (i) regularity of the metric components, (ii) regularity of the Christoffel components, and (iii) regularity of the curvature. 
We shall then embed the Kazakov--Solodukhin spacetime in a more general framework where these notions are clearly and cleanly separated. 
Finally we analyze aspects of the classical physics of these ``quantum deformed Schwarzschild spacetimes''. We shall discuss the surface gravity, the classical energy conditions, null and timelike geodesics, and the appropriate variant of Regge--Wheeler equation.

\bigskip
\noindent
{\sc Date:} Thursday 4 February 2021; \LaTeX-ed \today


}

%========================================================
\maketitle
%========================================================
%=======================================
\def\H{{\scriptscriptstyle{\mathrm{H}}}}
\def\ISCO{{\scriptscriptstyle{\mathrm{ISCO}}}}
\def\d{{\mathrm{d}}}
\def\O{{\mathcal{O}}}
\def\sign{{\mathrm{sign}}}

\clearpage
%=======================================
\section{Introduction}\label{S:intro}
%=======================================

The unification of general relativity and quantum mechanics is of the utmost importance in reconciling many open problems in theoretical physics today. 
One avenue of exploration towards a fully quantised theory of gravity is to, on a case--by--case basis, apply various quantum corrections to existing black hole solutions to the Einstein equations, and thoroughly analyse the resulting geometries through the lens of standard general relativity.
As with the majority of theoretical analysis, to make progress one begins by applying quantum--corrections to the simplest case; the Schwarzschild solution~\cite{Kazakov:1993}. 

Historically, various treatments of a quantum--corrected Schwarzschild metric have been performed in multiple different settings~\cite{Ali:2015, Calmet:2017, Md.:2018, Qi:2019, Shahjalal:2019, Eslamzadeh:2020, Good:2020, Nozari:2020, Nozari:2021}. 
A specific example of such a metric is the ``quantum deformed Schwarzschild metric'' derived by Kazakov and Solodukhin in reference~\cite{Kazakov:1993}. 
Much of the literature sees the original metric exported from the context of static, spherical symmetry into something dynamical, or else it invokes a different treatment of the quantum--correcting process to that performed in~\cite{Kazakov:1993} (see,~\emph{e.g.}, reference~\cite{Burger:2018}). 

The metric derived in reference~\cite{Kazakov:1993} invokes the following change to the line element for Schwarzschild spacetime in standard curvature coordinates:
\begin{eqnarray}
1-\frac{2m}{r} &\quad\longrightarrow\quad& \sqrt{1-\frac{a^{2}}{r^{2}}}-\frac{2m}{r},
\end{eqnarray}
so that
\begin{eqnarray}
\d s^{2} &=& -\left(\sqrt{1-\frac{a^{2}}{r^{2}}}-\frac{2m}{r}\right)\d t^{2} + \frac{\d r^{2}}{\sqrt{1-\frac{a^{2}}{r^{2}}}-\frac{2m}{r}} + r^{2}\,\d\Omega^{2}_{2}.
\label{k-s-metric}
\end{eqnarray}
To keep the metric components real, the $r$ coordinate must be  restricted to the range $r\in[a,\infty)$. So the ``centre'' of the spacetime at $r\to a$ is now a 2-sphere of finite area $A=4\pi a^2$.  The fact that the ``centre'' has now been ``smeared out'' to finite $r$ was originally hoped to render the spacetime regular.

\enlargethispage{40pt}
This metric was originally derived via an action principle which has its roots in the 2-D, (more precisely  (1+1)-D),  dilaton theory of gravity~\cite{Kazakov:1993, Russo:1992yg}:
\begin{equation}
S = -\frac{1}{8} \int \d^2 z \sqrt{-g} \left[ r^2 R^{(2)} - 2 (\nabla r)^2 + \frac{2}{\kappa} \, U(r) \right].
\label{k-s-action}
\end{equation}
Here \( R^{(2)} \) is the two--dimensional Ricci scalar, \( \kappa \) is a constant with dimensions of length, and \( U(r) \) is the ``dilaton potential''.

\clearpage
The action \eqref{k-s-action} yields two equations of motion, one of which is then used to derive the general form of the metric:
\begin{equation}
\d s^2 = -f(r)\;\d t^2 + \frac{\d r^2}{f(r)} + r^2 \;\d\Omega^2_2, \qquad \qquad
f(r) = -\frac{2m}{r} + \frac{1}{r}\int^r U(\rho)\,\d\rho.
\end{equation}
The dilaton potential \( U(r) \) is quantised within the context of the \( D=2 \) \( \sigma \)-model~\cite{Kazakov:1993, Russo:1992yg}, resulting in the specific metric \eqref{k-s-metric}.
Specifically, Kazakov and Solodukhin choose
\begin{equation}
U(r) = {r\over\sqrt{r^2-a^2}}. 
\end{equation}
Note that generic metrics of the form 
\begin{equation}
\d s^2 = -f(r)\;\d t^2 + \frac{\d r^2}{f(r)} + r^2 \;\d\Omega^2_2,
\end{equation}
where one does not necessarily make further assumptions about the function $f(r)$, have a long and complex history~\cite{Ted,Kiselev0,Kiselev1,Kiselev2}.

In Kazakov and Solodukhin's original work~\cite{Kazakov:1993}, they claim the metric \eqref{k-s-metric} is ``regular''. 
However, by this they just mean ``regular'' in the sense of the metric components (in this specific coordinate chart) being finite for all $r\in[a,\infty)$.
This is not the meaning of the word ``regular''  that is usually adopted in the GR community. 
We find it useful to carefully distinguish (i) regularity of the metric components, (ii) regularity of the Christoffel components, and (iii) regularity of the curvature. 
Indeed, within the GR community, the term ``regular'' means that the spacetime entirely is free of curvature singularities~\cite{Bardeen:1968, Roman:1983, Borde:1996, Bronnikov:2000, Moreno:2002, Ayon-Beato:2004, Hayward:2005, Bronnikov:2005, Bronnikov:2006, Lemos:2007, Ansoldi:2008, Lemos:2011, Bronnikov:2012, Bambi:2013, Bardeen:2014, Frolov:2014-a, Frolov:2014-b, Balart:2014, De-Lorenzo:2014, Frolov:2016, Fan:2016, Frolov:2017-a, Frolov:2017-b, Cano:2018, Bardeen:2018, Carballo-Rubio:2018-a, Carballo-Rubio:2018-b, Carballo-Rubio:2019-a, Carballo-Rubio:2019-b, Carballo-Rubio:2021}, with infinities in the curvature invariants being used as the typical diagnostic. 
While the metric \eqref{k-s-metric} is regular in terms of the metric components, it fails to be regular in terms of the Christoffel components, and has a Ricci scalar which is manifestly singular at \( r = a \):
\begin{equation}
R = \frac{2}{r^2} - \frac{2r^2-3a^2}{r (r^2-a^2)^{3\over2}}\,
 = {a\over(2a)^{3\over2} \, (r-a)^{3\over2}} - \frac{23}{4\left(2a\right)^{\frac{3}{2}}(r-a)^{\frac{1}{2}}} + \O(1).
\end{equation}

\enlargethispage{40pt}
The specific metric \eqref{k-s-metric} derived by Kazakov and Solodukhin falls in to a more general class of metrics given by
\begin{equation}
\d s_n^2 = -f_n(r) \d t^2 + \frac{\d r^2}{f_n(r)} + r^2\big(\d\theta^2 + \sin^2\theta \,\d\phi^2\big),
\label{E:g_n}
\end{equation}
where now we take
\begin{equation}
f_n(r) =   
\left(1-\frac{a^{2}}{r^{2}}\right)^{n\over2}-\frac{2m}{r}.
\label{E:f_n}
\end{equation}

Here \( n \in \{0\}\cup\{1,3,5,\dots\} \), \( r\in[a,\infty) \), and \( a\in(0,\infty) \). 
(Note, we include \( n = 0 \) as a special case since this reduces the metric to the Schwarzschild metric in standard curvature coordinates, which is useful for consistency checks).
We only consider odd values for \( n \) (excluding the \( n=0 \) Schwarzschild solution) as any even value of \( n \) will allow for the \( r \)-coordinate to continue down to \( r=0 \), and so produce a black-hole spacetime which is not regular at its core and hence not of interest in this work.

\enlargethispage{10pt}
The class of metrics described by equations~\eqref{E:g_n}--\eqref{E:f_n} has the following regularity structure:
\vspace{-15pt}
    \begin{itemize}\itemsep0pt
        \item \( n = 0 \) (Schwarzschild): Not regular;
        \item \( n \geq 1 \): Metric--regular;
        \item \( n \geq 3 \): Christoffel--symbol--regular;
        \item \( n \geq 5 \): Curvature--regular.
    \end{itemize}
\vspace{-10pt}
We wish to stress that, unlike reference~\cite{Kazakov:1993}, we make no attempt to \emph{derive} the class of metrics described by equations~\eqref{E:g_n}--\eqref{E:f_n}  from a modified action principle in this current work.
We feel that there are a number of technical issues requiring clarification in  the derivation presented in reference~\cite{Kazakov:1993}, so instead, we shall simply use the results of Kazakov and Solodukhin's work as inspiration and motivation for the analysis of our general class of metrics.
As such, our extended class of Kazakov--Solodukhin models can be viewed as another set of ``black hole mimickers''~\cite{Cardoso:2016, Visser:2009-a, Visser:2009-b, Visser:2003, Barcelo:2009, Simpson:2018, Simpson:2019-a, Lobo:2020, Simpson:2019-b, Berry:2020-a, Berry:2020-b, Boonserm:2018}, arbitrarily closely approximating standard Schwarzschild black holes, and so potentially of interest to observational astronomers~\cite{Barausse:2020}.

%=======================================
\section{Geometric analysis}
%=======================================

In this section we shall analyse the metric \eqref{E:g_n}, its associated Christoffel symbols,  and the various curvature tensor quantities derived therefrom.
 %
 
%=======================================
\subsection{Metric components}
%=======================================
 
We immediately enforce $a\neq 0$ since $a=0$ is trivially Schwarzschild, and in fact we shall specify $a>0$ since $a$ is typically to be identified with the Planck scale. 
At large $r$ and/or small $a$ we have:
\begin{equation}
f_n(r)= \left(1-\frac{a^{2}}{r^{2}}\right)^{n\over2}-\frac{2m}{r} = 1-\frac{2m}{r} - \frac{na^{2}}{2r^{2}} 
+ \O\left(\frac{a^4}{r^{4}}\right).
\end{equation} 
So the spacetime is asymptotically flat with mass $m$ for any fixed finite value of \( n \).
As $r\to a$ we note that for $n\geq1$ we have the finite limit
\begin{equation}
\lim_{r\to a} f_n(r) = - {2m\over a}.
\end{equation}
This is enough to imply metric--regularity.
Note however that for the radial derivative we have
\begin{equation}
f_n'(r) = {na^2\over r^3} \left(1-\frac{a^{2}}{r^{2}}\right)^{{n\over2}-1} +{2m\over r^2},
\end{equation}
and that only for $n\geq3$ do we have a finite limit
\begin{equation}
\lim_{r\to a} f_n'(r) =  {2m\over a^2}.
\end{equation}
Similarly for the second radial derivative
\begin{equation}
f_n''(r) = {na^2(na^2+a^2-3r^2)\over r^6} \left(1-\frac{a^{2}}{r^{2}}\right)^{{n\over2} -2} -{4m\over r^3},
\end{equation}
and only for $n\geq5$ do we have a finite limit
\begin{equation}
\lim_{r\to a} f_n''(r) =  -{4m\over a^3}.
\end{equation}
This ultimately is why we need $n\geq 3$ to make the Christoffel symbols regular, and $n\geq 5$ to make the curvature tensors regular.

%=======================================
\subsection{Event horizons}
%=======================================

Event horizons (Killing horizons) may be located by solving \( g_{tt}(r) = f_n(r) = 0 \), and so are implicitly characterized by 
\begin{equation}
r_H = 2m \left( 1 -{a^2\over r_H^2}\right)^{-{n\over2}}.
\label{E:horizon}
\end{equation}
This is not algebraically solvable for general \( n \), though we do have the obvious bounds that $r_H >2m$ and $r_H>a$. 

\enlargethispage{40pt}
Furthermore, for small \( a \) we can use (\ref{E:horizon}) to find an approximate horizon location 
by iterating the lowest-order approximation $r_H = 2m + \O(a^2/m)$ to yield
\begin{equation}
r_H = 2m \left\{ 1 +  {n a^2\over 8 m^2} + \O\left(a^4\over m^4\right) \right\}.
\end{equation}

Iterating a second time
\begin{equation}
r_H = 2m \left\{ 1 +  {n a^2\over 8 m^2} - {n(3n-2) a^4\over 128 m^4}   +\O\left({a^6\over m^6}\right) \right\}.
\end{equation}
We shall soon find that taking this second iteration is useful when estimating the surface gravity.
As usual, while event horizons are mathematically easy to work with, one should bear in mind that they are impractical for observational astronomers to deal with --- any physical observer limited to working in a finite region of space+time can at best detect apparent horizons or trapping horizons~\cite{observability}, see also reference~\cite{Hawking:2014}. In view of this intrinsic limitation, approximately locating the position of the horizon is good enough for all practical purposes.

%=======================================
\subsection{Christoffel symbols of the second kind}
%=======================================

Up to the usual symmetries, the non-trivial non-zero coordinate components of the Christoffel connection in this coordinate system are:
\begin{eqnarray}
\Gamma^t{}_{tr} &=&  -\Gamma^r{}_{rr}     =
{ 2m/r + n (a^2/r^2)(1-a^2/r^2)^{{n\over2} -1}
\over
2r  \{ (1-a^2/r^2)^{{n\over2}} - 2m/r \}
};
\nonumber\\
\Gamma^r{}_{tt} &=&
{ \{ 2m/r + n (a^2/r^2) (1-a^2/r^2)^{{n\over2} -1} \} \{(1-a^2/r^2)^{{n\over2}} - 2m/r \}
\over
2 r};
\nonumber\\
\Gamma^r{}_{\theta\theta} &=&  {\Gamma^r{}_{\phi\phi}\over \sin^2\theta}     =
2m - r(1-a^2/r^2)^{n\over2}.
\end{eqnarray}

The trivial non-zero components are 
\begin{eqnarray}
\Gamma^\theta{}_{r\theta} &=&  \Gamma^\phi{}_{r\phi}     = {1\over r};
\nonumber\\
\Gamma^\theta{}_{\phi\phi} &=& - \sin\theta\;\cos\theta;
\nonumber\\
\Gamma^\phi{}_{\theta\phi} &=&  \cot\theta.
\end{eqnarray}

Inspection of the numerators of \( \Gamma^{t}{}_{tr} \), \( \Gamma^{r}{}_{rr} \), and 
\( \Gamma^{r}{}_{\theta\theta} \) shows that (in this coordinate system) the Christoffel symbols are finite at \( r=a \) so long as \( n \geq 3 \).
Indeed as $r\to a$ we see
\begin{eqnarray}
\Gamma^t{}_{tr} &=&  -\Gamma^r{}_{rr}     \to -{1\over2a};
\qquad
\Gamma^r{}_{tt} \to -{2m^2\over a^3};
\nonumber\\
\Gamma^r{}_{\theta\theta} &=&  {\Gamma^r{}_{\phi\phi}\over \sin^2\theta}  \to2m ;
\quad\quad\;\;
\Gamma^\theta{}_{r\theta} \to  \Gamma^\phi{}_{r\phi}     = {1\over a}.
\end{eqnarray}

%=======================================
\subsection{Orthonormal components}
%=======================================

When a metric $g_{ab}$ is diagonal then the quickest way of calculating the orthonormal components of the Riemann and Weyl tensors is to simply set
\begin{equation}
R_{\hat a\hat b\hat c\hat d} = {R_{abcd}\over |g_{ac}|\,|g_{bd}|}; 
\qquad\qquad\qquad
C_{\hat a\hat b\hat c\hat d} = {C_{abcd}\over |g_{ac}|\,|g_{bd}|}.
\end{equation}
When a metric $g_{ab}$ is diagonal and a tensor $X_{ab}$ is diagonal then the quickest way of calculating the orthonormal components is to simply set
\begin{equation}
X_{\hat a\hat b} = {X_{ab}\over |g_{ab}|}.
\end{equation}
In both situations some delicacy is called for when crossing any horizon that might be present. 
Let us (using $-+++$ signature and assuming a diagonal metric) define
\begin{equation}
S = \sign(-g_{tt}) = \sign(g_{rr}).
\end{equation}
Then $S=+1$ in the domain of outer communication (above the horizon) and $S=-1$ below the horizon.

%=======================================
\subsection{Riemann tensor}
%=======================================

We shall now analyse what values of \( n \) result in non-singular components of various curvature tensors in an orthonormal basis \( (\t, \r, \th, \ph) \).
First, the non-zero orthonormal components of the Riemann tensor are:
\begin{align}
R_{\r\t\r\t} &= -\frac{2m}{r^3} - \frac{n a^2 \big[3 - (n+1)a^2/r^2\big](1-a^2/r^2)^{{n\over2} -2} }{2r^4},		\notag \\
R_{\r\th\r\th} &= R_{\r\ph\r\ph} = - R_{\th\t\th\t} = - R_{\ph\t\ph\t} =
-S\left\{ \frac{m}{r^3} + \frac{na^2 (1-a^2/r^2)^{{n\over2} -1} }{2r^{4}}\right\},	\notag \\
R_{\th\ph\th\ph} &=  \frac{2m}{r^3} + \frac{1-(1-a^2/r^2)^{n\over2}}{r^2}.
\end{align}

\enlargethispage{40pt}
Analysis of the numerator of \( R_{\r\t\r\t} \) shows that all of the orthonormal components of the Riemann tensor remain finite at \( r=a \) if and only if \( n\geq5 \).
Indeed as $r\to a$ (where $S\to-1$) we see
\begin{align}
R_{\r\t\r\t} &\to -\frac{2m}{a^3}; \qquad\qquad R_{\th\ph\th\ph} \to \frac{1}{a^2} + \frac{2m}{a^3}.
	\notag \\[5pt]
R_{\r\th\r\th} &= R_{\r\ph\r\ph} = - R_{\th\t\th\t} = - R_{\ph\t\ph\t} \to
+\frac{m}{a^3}.
\end{align}

Conversely at large $r$ (where  $S\to+1$) we see
\begin{align}
R_{\r\t\r\t} &= -\frac{2m}{r^3} + \O(a^2/r^4),		
\notag \\
R_{\r\th\r\th} &= R_{\r\ph\r\ph} = - R_{\th\t\th\t} = - R_{\ph\t\ph\t} =
-\frac{m}{r^3} +\O(a^2/r^4),	\notag \\
R_{\th\ph\th\ph} &=  \frac{2m}{r^3} +\O(a^2/r^4).
\end{align}
So, as it should, the spacetime curvature asymptotically approaches that of Schwarzschild.

%=======================================
\subsection{Ricci tensor}
%=======================================
The non-zero orthonormal components of the Ricci tensor are:
\begin{align}
R_{\t\t} &= - R_{\r\r} = - S \; \frac{na^2}{2r^4} \big[ 1 - (n-1)a^2/r^2 \big] (1-a^2/r^2)^{{n\over2} -2},	\notag \\
R_{\th\th} &= R_{\ph\ph} = \frac{1}{r^2} - \frac{1}{r^{2}} \big[1 + (n-1)a^2/r^2 \big] (1-a^2/r^2)^{{n\over2} -1}.
\end{align}

Analysis of the \( R_{\r\r} \) component shows that all of the components of the Ricci tensor remain finite at \( r=a \) so long as \( n\geq5 \).
Indeed as $r\to a$ we see
\begin{align}
R_{\t\t} &=  - R_{\r\r} \to 0,	
\qquad\qquad
R_{\th\th} = R_{\ph\ph} \to \frac{1}{a^2} .
\end{align}

Conversely at large $r$ we have
\begin{align}
R_{\t\t} &= - R_{\r\r} =R_{\th\th} = R_{\ph\ph}= -\frac{na^2}{2r^4} 
+\O(a^4/r^6).
\end{align}

%=======================================
\subsection{Ricci scalar}
%=======================================

As stated in Section \ref{S:intro}, our class of metrics is only curvature regular for \( n\geq5 \), where \( n \) is an odd integer.
Indeed, in general we have
\begin{equation}
R = \frac{2}{r^2} - (1-a^2/r^2)^{{n\over2}-2}  \bigg\{\frac{2 + (n-4)a^2/r^2 + (n-2)(n-1)a^4/r^4}{ r^{2} }\bigg\},
\end{equation}
and so the spacetime is non-singular at \( r=a \) if and only if \( n\geq5\).
Furthermore, any \( n\geq5 \) spacetime has positive scalar curvature at \( r=a \), where
$R \to  \frac{2}{a^2}$.

As an explicit example, 
\begin{equation}
R_{n=5} = \frac{2}{r^2} - \sqrt{r^2-a^2} \, \bigg\{\frac{2r^4 + a^2r^2 + 12a^4}{r^7}\bigg\},
\end{equation}
which is indeed singularity--free in the region \( r \in [a,\infty) \) and positive at \( r=a \).

%=======================================
\subsection{Einstein tensor}
%=======================================

The non-zero components of the Einstein tensor are
\begin{align}
G_{\t\t} &= - G_{\r\r} = \frac{S}{r^2}\left\{ 1
 - \left[1+(n-1){a^2\over r^2}\right] \left(1-{a^2\over r^2}\right)^{(n-2)/2}\right\},	\notag \\
G_{\th\th} &= G_{\ph\ph} = - \frac{na^2}{2r^4} \left[1-(n-1){a^2\over r^2}\right]
\left(1-{a^2\over r^2}\right)^{(n-4)/2}.
\end{align}

Analysis of the \( G_{\th\th} \) component reveals that the Einstein tensor remains finite in all of its orthonormal components if and only if \( n\geq5 \).
Indeed as $r\to a$ (where $S\to-1$) we see
\begin{align}
G_{\t\t} &=  - G_{\r\r} \to  - \frac{1}{a^2} ,	
\qquad\qquad
G_{\th\th} = G_{\ph\ph} \to 0.
\end{align}
At large $r$ (where $S\to+1$)  we have
\begin{equation}
G_{\t\t} =  - G_{\r\r} = G_{\th\th} = G_{\ph\ph} = -{na^2\over2r^4} + \O(a^4/r^6).
\end{equation}

%=======================================
\subsection{Weyl tensor}
%=======================================

The non-zero components of the Weyl tensor are
\begin{align}
C_{\r\t\r\t} &= 2S\, C_{\r\th\r\th} = 2 S\, C_{\r\ph\r\ph} 
= -2 S\,C_{\th\t\th\t} = -2 S\, C_{\ph\t\ph\t} = - C_{\th\ph\th\ph} 	\notag \\
&= - \frac{2m}{r^3} + {(1-a^2/r^2)^{{n\over2} -2} -1\over 3r^2}
\notag \\
&\qquad - a^2(1-a^2/r^2)^{{n\over2} -2} 
\left\{\frac{ (5n+4) - (n+2)(n+1)a^2/r^2}{6r^{4}}\right\}.
\end{align}
Thus, the components of the Weyl tensor remain finite at \( r=a \) so long as \( n\geq5 \).

Indeed as $r\to a$ (where $S\to-1$) we see
\begin{align}
C_{\r\t\r\t} &= -2C_{\r\th\r\th} = -2C_{\r\ph\r\ph} = +2 C_{\th\t\th\t} 
= +2 C_{\ph\t\ph\t} = - C_{\th\ph\th\ph}  \to - {1\over 3a^2} - {2m\over a^3}. 
\end{align}
At large $r$ (where $S\to+1$) we find
\begin{align}
C_{\r\t\r\t} &= 2C_{\r\th\r\th} = 2C_{\r\ph\r\ph} = -2 C_{\th\t\th\t} = -2 C_{\ph\t\ph\t} = - C_{\th\ph\th\ph}  = - {2m\over r^3} -{na^2\over r^4} +\O(a^4/r^6). 
\end{align}

%=======================================
\subsection{Weyl scalar}
%=======================================
The Weyl scalar is defined by $C_{abcd}\, C^{abcd}$. In view of all the symmetries of the spacetime one can show that  $C_{abcd} \, C^{abcd} = 12 (C_{\r\t\r\t})^2$, so one gains no additional behaviour beyond looking at the Weyl tensor itself.
Thus, for purposes of tractability we will only display the result for \( n=5 \) at \( r=a \) in order to show that the \( n=5 \) spacetime is indeed regular at \( r=a \):
\begin{equation}
\left. (C_{abcd}\, C^{abcd})_{n=5}\right|_{r=a} = \frac{4(6m+a)^2}{3a^6}.
\end{equation}

%=======================================
\subsection{Kretschmann scalar}
%=======================================

The Kretschmann scalar is given by
\begin{equation}
K = R_{abcd}R^{abcd} = C_{abcd}C^{abcd} + 2R_{ab}R^{ab} - \frac{1}{3} R^2.
\end{equation}
The general result is rather messy and does not provide much additional insight into the spacetime.
Thus, for purposes of tractability we will only display the result for \( n=5 \) at \( r=a \) in order to show that the \( n=5 \) spacetime is indeed regular at \( r=a \):
\begin{equation}
\left. K_{n=5}\right|_{r=a} = \frac{4}{a^6} \big(a^2+4am+12m^2\big).
\end{equation}
The fact that the Kretschmann scalar is positive definite, and can be written as a sum of squares, is ultimately a due to spherical symmetry and the existence of a hypersurface orthogonal Killing vector~\cite{novel}.

%=================================================
\section{Surface gravity and Hawking temperature}\label{kappa}
%=================================================

Let us calculate the surface gravity at the event horizon for the generalised QMS spacetime. 
Because we are working in curvature coordinates we always have~\cite{DBH1}
\begin{equation}
\kappa_H = \lim_{r\to r_H} {1\over2} {\partial_r g_{tt}\over\sqrt{g_{tt}\; g_{rr}}}.
\end{equation}
Thence
\begin{equation}
\kappa_H = \left.{1\over2}\; \partial_r f_n(r)\right|_{r_H} 
= {m\over r_H^2} + {na^2\over 2 r_H^3} \left(1-{a^2\over r_H^2}\right)^{{n\over2}-1}.
\end{equation}
Using equation (\ref{E:horizon}) we can also rewrite this as
\begin{equation}
\kappa_H = 
 {m\over r_H^2} \left\{1 + {na^2\over r_H^2-a^2} \right\} .
\end{equation}
This result is, so far, exact.
Given that the horizon location is not analytically known for general $n$, we shall use the asymptotic result $r_{H}= 2m \left\{1+ \frac{n a^2}{8m^2}  - {n(3n-2) a^4\over 128 m^4} + \mathcal{O}({a^6\over m^6})\right\}$. Thence
\begin{equation}
\kappa_H = 
 {1\over 4m}\left\{ 1 - {n(n-1)a^4\over32 m^4}  + \O(a^6/m^6)\right\}. 
\end{equation}
Note the potential $\O(a^2/m^2)$ term vanishes (which is why we estimated $r_H$ up to $\O(a^4)$).
As usual the Hawking temperature is simply $k_B T_H = {1\over2\pi} \,\hbar \,\kappa_H$.

%===========================================
\section{Stress-energy tensor}\label{stress}
%===========================================

Let us examine the Einstein field equations for this spacetime. 
Above the horizon, for $r > r_H$,  we have 
\begin{equation}
8\pi\,\rho = G_{\t\t};
\qquad\qquad
8\pi\,p_r = G_{\r\r}.
\end{equation}
Below the horizon, for $r < r_H$,  we have 
\begin{equation}
8\pi\,\rho = G_{\r\r};
\qquad\qquad
8\pi\,p_r = G_{\t\t}.
\end{equation}
But then regardless of whether one is above or below the horizon one has
\begin{align}
\rho &= - p_r= \frac{1}{8\pi r^2}\left\{ 1
 - \left[1+(n-1){a^2\over r^2}\right] \left(1-{a^2\over r^2}\right)^{(n-2)/2}\right\},	\notag \\
p_\perp&= - \frac{na^2}{16\pi r^4} \left[1-(n-1){a^2\over r^2}\right]
\left(1-{a^2\over r^2}\right)^{(n-4)/2}.
\end{align}

By inspection, for $n>1$ we see that $p_\perp(r)=0$ at $r= \sqrt{n-1} \;a$. Indeed we see that $p_\perp(r)>0$ for $r< \sqrt{n-1} \;a$ and $p_\perp(r)<0$ for $r> \sqrt{n-1} \;a$.
The analagous result for $\rho(r)$ is not analytically tractable (though it presents no numerical difficulty) as by inspection it amounts to finding the roots of 
\begin{equation}
(r^2-a^2) r^n - (r^2-a^2)^{n\over2} (r^2+ (n-1)a^2) = 0.
\end{equation}
We note that asymptotically
\begin{equation}
\label{E:rho}
\rho = - {na^2\over 16\pi r^4} + \O(a^4/ r^6),
\end{equation}
and 
\begin{equation}
p_\perp = - {na^2\over 16\pi r^4} + \O(a^4/ r^6).
\end{equation}

An initially surprising result is that the stress-energy tensor has no dependence on the mass \( m \) of the spacetime.
To see what is going on here, consider the Misner--Sharp quasi-local mass
\begin{equation}
1-{2m(r)\over r} = f_n(r) \qquad\implies \qquad 
m(r) = m +{r\over2} \left\{ 1- \left(1-{a^2\over r^2}\right)^{n\over2} \right\}.
\end{equation}
Then, noting that $m = m(r) + 4\pi \int_r^\infty \rho(\bar r) \bar r^2 \d\bar r$ above the horizon, we see
\begin{equation}
4\pi \int_r^\infty \rho(\bar r) \bar r^2 \d\bar r = -{r\over2} \left\{ 1- \left(1-{a^2\over r^2}\right)^{n\over2} \right\}.
\end{equation}
Here the RHS is manifestly independent of $m$. 
Consequently, without need of any detailed calculation, $\rho(r)$ is manifestly independent of $m$. 
As an aside note that $m(r_H) = {r_H\over2}$, so we could also write
$ m(r) = {r_H\over2} + 4\pi \int_{r_H}^r \rho(\bar r) \bar r^2 \d\bar r$.

%===========================================
\section{Energy conditions}\label{S:ecs}
%===========================================

The classical energy conditions are constraints on the stress-energy tensor that attempt to keep various aspects of ``unusual physics'' under control~\cite{ECs, Tipler:1978, Borde:1987, Klinkhammer:1991, Ford:1994, book, Fewster:2002, twilight, Visser:1997, Visser:1999, Visser:1996, Roman:2004, Cattoen:2006, Visser:1997-b, Fewster:2010, Zaslavskii:2010, LNP-survey, Martin-Moruno:2013-a, Martin-Moruno:2013-b, Curiel:2014, Martin-Moruno:2015}. While it can be argued that the classical energy conditions are not truly fundamental~\cite{twilight, Visser:1996, LNP-survey}, often being violated by semi-classical quantum effects, they are nevertheless extremely useful indicative probes, well worth the effort required to analyze them. 

%========================================================
\subsection{Null energy condition}
%========================================================

A necessary and sufficient condition for the null energy condition (NEC) to hold is that both $\rho + p_{r}\geq 0$ and $\rho+p_\perp\geq 0$  for all $r$, $a$, $m$.  
Since $\rho=-p_{r}$, the former inequality is trivially satisfied, and for all $r\geq a$ we may simply consider
\begin{equation}
\rho+p_\perp  = \frac{1}{8\pi r^2} \left\{ 1- \frac{(1-a^2/r^2)^{{n\over2}-2}}{2}
 \bigg[ 2 + (3n-4)a^2/r^2 - (n+2)(n-1)a^4/r^4 \bigg] \right\}.
\end{equation}
Whether or not this satisfies the NEC depends on the value for \( n \).
Furthermore, for no value of \( n \) is the NEC \emph{globally} satisfied.

\begin{figure}[!htbp]
\begin{center}
\includegraphics[scale=0.65]{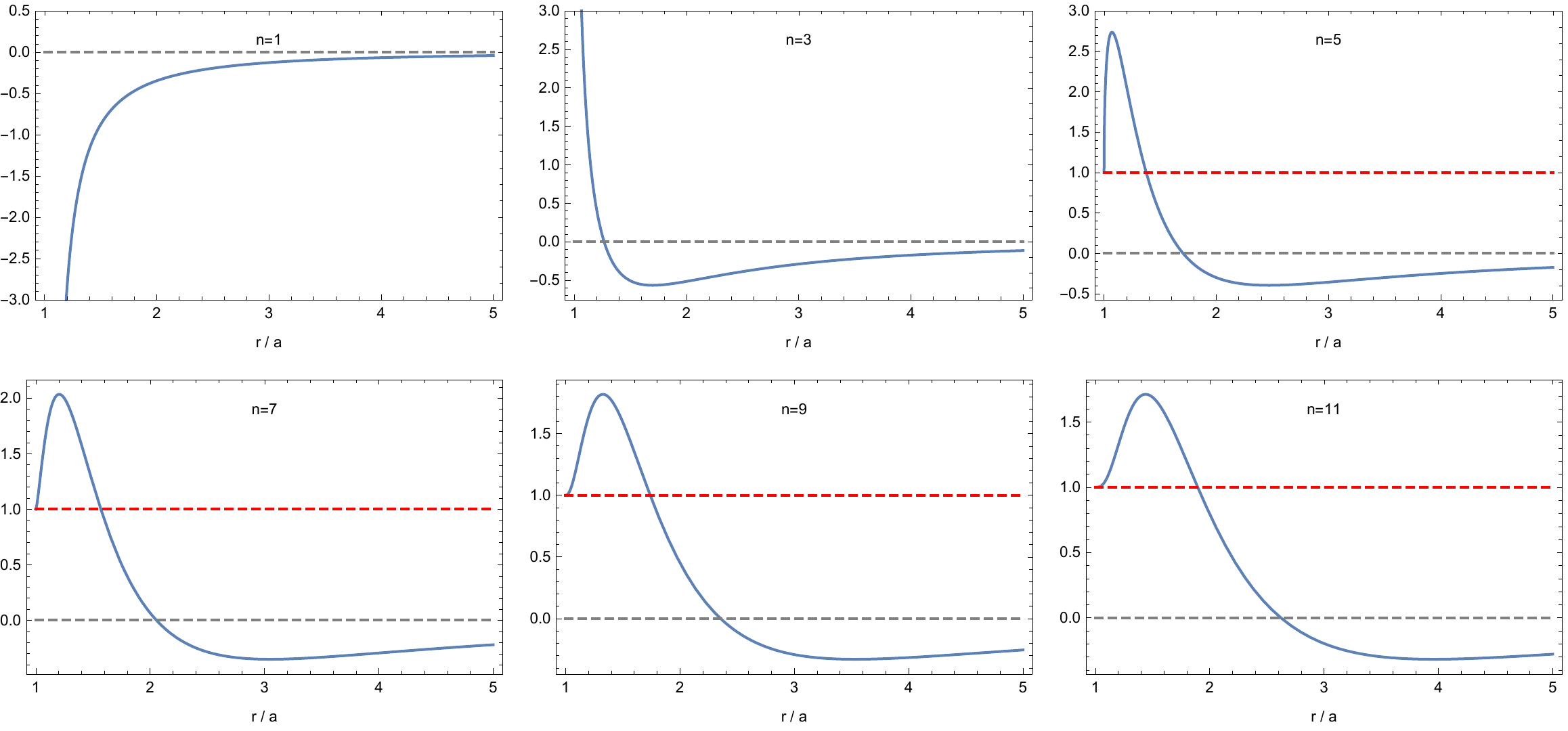}
\caption{Plots of the NEC for several values of $n$. Here the $y$-axis depicts $8\pi r^2(\rho + p_{\perp})$, plotting against $r/a$ on the $x$-axis. Of particular interest are the qualitative differences in behaviour as $r/a\rightarrow 1$; we see divergent behaviour for the $n=1$ and $n=3$ cases, whilst for $n\geq 5$ we see $8\pi r^2(\rho + p_{\perp})\rightarrow 1$ as $r\rightarrow a$. This is ultimately due to the fact that the $n\geq 5$ cases are \emph{curvature regular}, with globally finite stress-energy components.}
\label{F:nec}
\end{center}
\end{figure}

Provided $n\geq 5$, so that the limits exist, we have
\begin{equation}
\lim_{r\to a} (\rho+p_\perp) = +{1\over8\pi a^2}.
\end{equation}
So the NEC is definitely satisfied deep in the core of the system.
Note that at asymptotically large distances
\begin{equation}
\rho+p_t  = - {na^2\over 8\pi r^4} + \O(a^4/ r^6).
\end{equation}
So the NEC (and consequently all the other classical point-wise energy conditions) are  always violated at asymptotically large distances.
However, for some values of \( n \), there are bounded regions of the spacetime in which the NEC is satisfied.
See Figure \ref{F:nec}.

%%=============================================================
%%=============================================================
%%=============================================================
%%=============================================================

%================================================
\subsection{Weak energy condition}
%================================================
In order to satisfy the weak energy condition (WEC) we require the NEC be satisfied, and  in addition \( \rho \geq 0 \).
But in view of the asymptotic estimate (\ref{E:rho}) for $\rho$ we see that the WEC is always violated at large distances.
Furthermore, it can be seen from Table~\ref{T:1} that the region in which the NEC is satisfied is always larger than that in which $\rho$  is positive (this would be as good as impossible to prove analytically for general \( n \)).
Thus, we can conclude (see Table \ref{T:2})  that the WEC is satisfied for smaller regions than the NEC for all values of \( n \).

%================================================
\subsection{Strong energy condition}
%================================================

In order to satisfy the strong energy condition (SEC) we require the NEC to be satisfied, and  in addition \( \rho + p_r + 2p_\perp = 2p_\perp \geq 0 \).
But regardless of whether one is above or below the horizon, the second of these conditions $p_\perp\geq0$ amounts to
\begin{equation}
0 < a < r \leq a \sqrt{n-1}.
\end{equation}
However, it can be seen from Table \ref{T:1} that the region in which the NEC is satisfied is always smaller than that in which $p_\perp$ is positive (this would be as good as impossible to prove analytically for general \( n \)).
Thus, we can conclude (see Table \ref{T:2}) that the SEC is satisfied in the same region as the NEC for all values of \( n \).

%================================================
\subsection{Dominant energy condition}
%================================================

The dominant energy condition is the strongest of the standard classical energy conditions.
Perhaps the best physical interpretation of the DEC is that for any observer with timelike 4-velocity $V^a$ the flux vector $F^a = T^{ab}\, V_b$ is non-spacelike (timelike or null). 
It is a standard result that in spherical symmetry (in fact for any type~I stress-energy tensor) this reduces to positivity of the energy density $\rho>0$ combined with the condition $|p_i| \leq \rho$. 
Since in the current framework for the radial pressure we always have $p_r = -\rho$, the only real constraint comes from demanding $|p_\perp| \leq \rho$.
But this means we want \emph{both} $\rho+p_\perp\geq 0$ \emph{and} $\rho-p_\perp\geq 0$.
The first of these conditions is just the NEC, so the only new constraint comes from the second condition. 
By inspection, it can be seen from Table \ref{T:1} that the region in which the NEC is satisfied is always larger than that in which $\rho-p_\perp $  is positive (this would be as good as impossible to prove analytically for general \( n \)).
Thus, we can conclude (see Table \ref{T:2})  that the DEC is satisfied for smaller regions than the NEC for all values of \( n \).

%%%%%%%%%%%%%%%%%%%%%%%%%%%%%%%%%%%%%%%%%%%%%%
%%%%%%%%%%%%%%%%%%%%% Table 1 %%%%%%%%%%%%%%%%%%%%%
%%%%%%%%%%%%%%%%%%%%%%%%%%%%%%%%%%%%%%%%%%%%%%
\begin{table}[!htbp]
\centering
    \caption{Regions of the spacetime where the orthonormal components of the stress-energy tensor satisfy certain inequalities.}
    \tabulinesep=1.5mm
    \begin{tabu}{|c|c|c|c|c|}
        \hline
         $n$ & $\rho+p_\perp\geq0$ & $\rho\geq0$ & $p_\perp\geq0$ & $\rho-p_\perp\geq0$ \\
        \hline \hline
         0 & $a<r<\infty$ & $a<r<\infty$ & $a<r<\infty$ & $a<r<\infty$ \\
        \hline
         1 & globally violated & globally violated & globally violated & globally violated \\
        \hline
         3 & $a<r \lessapprox 1.26595a$ & $a<r \lessapprox 1.07457a$ & $a<r \lessapprox 1.41421a$ & globally violated\\
        \hline
        5 & $a<r \lessapprox 1.70468a$ & $a<r \lessapprox 1.37005a$ & $a<r \leq 2a$ & $a<r \lessapprox 1.00961a$ \\
        \hline
        7 & $a<r \lessapprox 2.05561a$ & $a<r \lessapprox 1.62933a$ & $a<r \lessapprox 2.44949a$ & $a<r \lessapprox 1.11129a$ \\
        \hline
        9 & $a<r \lessapprox 2.35559a$ & $a<r \lessapprox 1.85537a$ & $a<r \lessapprox 2.82843$ & $a<r \lessapprox 1.23076a$ \\
        \hline
        11 & $a< r\lessapprox 2.62173a$ & $a<r \lessapprox 2.05757a$ & $a<r \lessapprox 3.16228a$ & $a<r \lessapprox 1.34552a$ \\
        \hline
        \vdots & \vdots & \vdots & \vdots & \vdots \\
        \hline
    \end{tabu}
\label{T:1}
\end{table}
%%%%%%%%%%%%%%%%%%%%%%%%%%%%%%%%%%%%%%%%%%%%%%
%%%%%%%%%%%%%%%%%%%%% Table 2 %%%%%%%%%%%%%%%%%%%%%
%%%%%%%%%%%%%%%%%%%%%%%%%%%%%%%%%%%%%%%%%%%%%%
\begin{table}[!htbp]
\centering
    \caption{Regions of the spacetime where the energy conditions are satisfied.}
    \tabulinesep=1.5mm
    \begin{tabu}{|c|c|c|c|c|}
        \hline
         $n$ & NEC & WEC & SEC & DEC \\
        \hline \hline
         0 & $a<r<\infty$ & $a<r<\infty$ & $a<r<\infty$ & $a<r<\infty$ \\
        \hline
         1 & globally violated & globally violated & globally violated & globally violated \\
        \hline
         3 & $a<r \lessapprox 1.26595a$ & $a<r \lessapprox 1.07457a$ & same as NEC & globally violated\\
        \hline
        5 & $a<r \lessapprox 1.70468a$ & $a<r \lessapprox 1.37005a$ & same as NEC & $a<r \lessapprox 1.00961a$ \\
        \hline
        7 & $a<r \lessapprox 2.05561a$ & $a<r \lessapprox 1.62933a$ & same as NEC & $a<r \lessapprox 1.11129a$ \\
        \hline
        9 & $a<r \lessapprox 2.35559a$ & $a<r \lessapprox 1.85537a$ & same as NEC & $a<r \lessapprox 1.23076a$ \\
        \hline
        11 & $a< r\lessapprox 2.62173a$ & $a<r \lessapprox 2.05757a$ & same as NEC & $a<r \lessapprox 1.34552a$ \\
        \hline
        \vdots & \vdots & \vdots & \vdots & \vdots \\
        \hline
    \end{tabu}
\label{T:2}
\end{table}
%%%%%%%%%%%%%%%%%%%%%%%%%%%%%%%%%%%%%%%%%%%%%%%
%%%%%%%%%%%%%%%%%%%%%%%%%%%%%%%%%%%%%%%%%%%%%%%

\newpage
%============================================
\section{ISCO and photon sphere analysis}\label{ISCO}
%============================================

We have the generalised quantum modified Schwarzschild metric

\begin{eqnarray}
\d s^{2} &=& -\left\{\left(1-\frac{a^{2}}{r^{2}}\right)^{\frac{n}{2}}-\frac{2m}{r}\right\}\d t^{2} + \frac{\d r^{2}}{\left(1-\frac{a^{2}}{r^{2}}\right)^{\frac{n}{2}}-\frac{2m}{r}} + r^{2}\,\d\Omega^{2}_{2}.
\label{k-s-metric-repeat}
\end{eqnarray}

Let us now find the location of both the photon sphere for massless particles, and the ISCO for massive particles, as functions of the parameters $m$, $n$, and $a$.
Consider the tangent vector to the worldline of a massive or massless particle, parameterized by some arbitrary affine parameter, $\lambda$: 
\begin{equation}
g_{ab}\frac{\d x^{a}}{\d\lambda}\frac{\d x^{b}}{\d\lambda}=-g_{tt}\left(\frac{\d t}{\d\lambda}\right)^{2}+g_{rr}\left(\frac{\d r}{\d\lambda}\right)^{2}+r^{2}\left\lbrace\left(\frac{\d\theta}{\d\lambda}\right)^{2}+\sin^{2}\theta \left(\frac{\d\phi}{\d\lambda}\right)^{2}\right\rbrace.
\end{equation}
We may, without loss of generality, separate the two physically interesting cases (timelike and null) by defining:
\begin{equation}
\epsilon = \left\{
\begin{array}{rl}
-1 & \qquad\mbox{massive particle, \emph{i.e.} timelike worldline} \\
0 & \qquad\mbox{massless particle, \emph{i.e.} null worldline}.
\end{array}\right. 
\end{equation}
That is, $\d s^{2}/\d\lambda^2=\epsilon$. Due to the metric being spherically symmetric we may fix $\theta=\frac{\pi}{2}$ arbitrarily and view the reduced equatorial problem:
\begin{equation}
g_{ab}\frac{\d x^{a}}{\d\lambda}\frac{\d x^{b}}{\d\lambda}=-g_{tt}\left(\frac{\d t}{\d\lambda}\right)^{2}+g_{rr}\left(\frac{\d r}{\d\lambda}\right)^{2}+r^{2}\left(\frac{\d\phi}{\d\lambda}\right)^{2}=\epsilon.
\end{equation}

The Killing symmetries yield the following expressions for the conserved energy $E$ and angular momentum $L$ per unit mass:
\begin{equation}
\left\{{\left(1-\frac{a^2}{r^2}\right)^{\frac{n}{2}}}-\frac{2m}{r}\right\}\left(\frac{\d t}{\d\lambda}\right)=E \ ; \qquad\quad r^{2}\left(\frac{\d\phi}{\d\lambda}\right)=L.
\end{equation}
Hence
\begin{equation}
\left\{{\left(1-\frac{a^2}{r^2}\right)^{\frac{n}{2}}}-\frac{2m}{r}\right\}^{-1}\left\lbrace -E^{2}+\left(\frac{\d r}{\d\lambda}\right)^{2}\right\rbrace+\frac{L^{2}}{r^{2}}=\epsilon,
\end{equation}
implying
\begin{equation}
\left(\frac{\d r}{\d\lambda}\right)^{2}=E^{2}+\left\{{\left(1-\frac{a^2}{r^2}\right)^{\frac{n}{2}}}
-\frac{2m}{r}\right\}\left\lbrace\epsilon-\frac{L^{2}}{r^{2}}\right\rbrace.
\end{equation}
This gives ``effective potentials" for geodesic orbits as follows:
\begin{equation}
V_{\epsilon}(r)=\left\{{\left(1-\frac{a^2}{r^2}\right)^{\frac{n}{2}}}-\frac{2m}{r}\right\}\left\lbrace -\epsilon+\frac{L^{2}}{r^{2}}\right\rbrace .
\end{equation}

%-------------------------------------------------------------
\subsection{Photon orbits}
%-------------------------------------------------------------
For a photon orbit we have the massless particle case $\epsilon=0$. Since we are in a spherically symmetric environment, solving for the locations of such orbits amounts to finding the coordinate location of the ``photon sphere''. These circular orbits occur at $V_{0}^{'}(r)=0$. That is:
\begin{equation}\label{V0r}
V_{0}(r)=\left\{\left(1-{a^2\over r^2}\right)^{\frac{n}{2}}-\frac{2m}{r}\right\}
\left\{\frac{L^{2}}{r^{2}}\right\},
\end{equation}
leading to:
\begin{equation}
    V_{0}^{'}(r) = \frac{L^2}{r^4}\left\lbrace 6m+r\left(1-\frac{a^2}{r^2}\right)^{\frac{n}{2}-1}\left[(n+2)\frac{a^2}{r^2}-2\right]\right\rbrace \ .
\end{equation}

Solving $V_{0}^{'}(r)=0$ analytically is intractable, but we may perform a Taylor series expansion of the above function about $a=0$ for a valid approximation (recall $a$ is associated with the Planck length). 

To fifth-order this yields:
\begin{equation}
    V_{0}^{'}(r) = \frac{2L^{2}}{r^4}(3m-r)+\frac{2L^{2}na^{2}}{r^5}-\frac{3na^{4}L^{2}(n-2)}{4r^{7}} + \mathcal{O}\left(L^{2}a^{6}/r^{9}\right) \ .
\end{equation}

Equating this to zero and solving for $r$ yields:

\begin{equation}\label{rphgenn}
    r_\gamma = 3m \left\{1 + \frac{a^2n}{(3m)^2} 
    - {n(11n-6)a^4\over 8(3m)^4} + \mathcal{O}(a^6/m^6) \right\}\ .
\end{equation}

The $a=0$, (or $n=0$), Schwarzschild sanity check reproduces $r_{\gamma}=3m$, the expected result.

To verify stability, we check the sign of $V_{0}^{''}(r)$:
\begin{equation}\label{V''}
    V_{0}^{''}(r) = -\frac{L^{2}}{r^{4}}\left\lbrace\frac{24m}{r}-\left(1-\frac{a^2}{r^2}\right)^{\frac{n}{2}-2}\left[6-(7n+12)\frac{a^{2}}{r^{2}}+(n+2)(n+3)\frac{a^4}{r^4}\right]\right\rbrace \ .
\end{equation}

We now substitute the approximate expression for $r_\gamma$ into Eq.~(\ref{V''}) to determine the sign of $V_{0}''(r_\gamma)$.
We find:
\begin{equation}
V_0''(r_\gamma) = -{2L^2\over 81 m^4} \left\{ 1-{3n a^2\over (3 m)^2} + {n(67n-6)a^4 \over 8 (3 m)^4} 
+ \O(a^6/m^6) \right\}
\end{equation}
Given that all bracketed terms to the right of the $1$ are strictly subdominant in view of $a\ll m$, we may conclude that $V_{0}^{''}(r_{\gamma})<0$, and hence the null orbits at $r=r_{\gamma}$ are unstable.

Let us now recall the generalised form of equation~(\ref{V0r}), and specialise to $n=5$ (the lowest value for $n$ for which our quantum deformed Schwarzschild spacetime is \emph{regular}). We have:
\begin{eqnarray}
    V_{0}(r,n=5) &=& \frac{L^2}{r^2}\left\lbrace\left(1-\frac{a^{2}}{r^2}\right)^{\frac{5}{2}}-\frac{2m}{r}\right\rbrace \ ; \\
    && \nonumber \\
    V_{0}'(r,n=5) &=& \frac{L^2}{r^4}\left\lbrace 6m -\sqrt{r^2-a^2}\left(2-\frac{9a^2}{r^2}+\frac{7a^4}{r^4}\right)\right\rbrace \ .
\end{eqnarray}
Once again setting this to zero and attempting to solve analytically is an intractable line of inquiry, and we instead inflict Taylor series expansions about $a=0$. 

To fifth-order we have the following
\begin{eqnarray}
    V_{0}'(r,n=5) &=& -\frac{2L^2}{r^3} \left\{1-{3m\over r} - \frac{5a^2}{r^2}
    + {45 L^2 a^4\over 8 r^4}+\mathcal{O}(a^6/r^6) \right\} \ ; \nonumber \\
    && \nonumber \\
    \Longrightarrow \quad r_{\gamma} &=& 3m\left\{ 1  + \frac{5a^2}{(3m)^2} -{245a^4\over 8(3 m)^4} + \mathcal{O}(a^6/m^6)\right\}  \ ,
\end{eqnarray}
which is consistent with the result for general $n$ displayed in Eq.~(\ref{rphgenn}).

%-------------------------------------------------------------
\subsection{ISCOs}
%-------------------------------------------------------------

For massive particles the geodesic orbit corresponds to a timelike worldline and we have the case that $\epsilon=-1$. Therefore:
\begin{equation}
V_{-1}(r) = \left\{\left(1-\frac{a^2}{r^2}\right)^{\frac{n}{2}}-\frac{2m}{r}\right\}
\left\{1+\frac{L^{2}}{r^{2}}\right\}  ,
\end{equation}
and it is easily verified that this leads to:
\begin{equation}
V_{-1}^{'}(r) = \frac{2m(3L^2+r^2)}{r^4} + \frac{(1-a^2/r^2)^{{\frac{n}{2}-1}}}{r^{3}}
\left[n a^2 +L^2\left((n+2)\frac{a^2}{r^2}-2\right)\right] \  .
\end{equation}

For small $a$ we have
\begin{equation}
V_{-1}(r) = \left\{1+{L^2\over r^2}\right\} \left\{1 -{2m\over r}-{na^2\over2r^2}
+{n(n-2) a^4\over 8 r^4} +\O\left(a^6\over r^6\right)\right\},
\end{equation}

and
\begin{equation}
V_{-1}^{'}(r) = {2(L^2(3m-r) + mr^2)\over r^4} +{(2L^2+r^2)n a^2\over r^5} 
- {(3L^2+2r^2)n(n-2)a^4\over 4r^7} 
+\O\left(a^6\over r^{{7}}\right).
\end{equation}

Equating this to zero and rearranging for $r$ presents an intractable line of inquiry. Instead it is preferable to assume a fixed circular orbit at some $r=r_{c}$, and rearrange the required angular momentum $L_{c}$ to be a function of $r_{c}$, $m$, and $a$. It then follows that the innermost circular orbit shall be the value of $r_{c}$ for which $L_{c}$ is minimised. It is of course completely equivalent to perform this procedure for the mathematical object $L_{c}^{2}$, and we do so for tractability.

Hence if $V_{-1}^{'}(r_{c})=0$, we have:
\begin{equation}\label{Lcsq}
    L_{c}^{2} = \frac{na^2\left(1-\frac{a^2}{r^2}\right)^{\frac{n}{2}}+2mr\left(1-\frac{a^2}{r^2}\right)}{\left(1-\frac{a^2}{r^2}\right)^{\frac{n}{2}}\left[2-(n+2)\frac{a^2}{r^2}\right]-\frac{6m}{r}\left(1-\frac{a^2}{r^2}\right)}.
\end{equation}

For small $a$ we have
\begin{equation}\label{Lcsqa0}
L_c^2 = {mr^2\over r-3m} + {n r (r-m)a^2\over2(r-3m)^2} 
-
{n\{(2n+4)r^2 +(5n-18) mr -9(n-2) m^2 \}a^4\over 8 r (r-3m)^3}
+ \O(a^6).
\end{equation}

As a consistency check, for large $r_{c}$ (\emph{i.e.} $r_{c}\gg a, m$) we observe from the dominant term of Eq.~(\ref{Lcsqa0}) that $L_{c}\approx\sqrt{mr_{c}}$, which is consistent with the expected value when considering circular orbits in weak-field GR. Indeed it is easy to check that for large $r$ we have $L_{c}^{2} = mr_{c} + \mathcal{O}(1)$. Note that in classical physics the angular momentum per unit mass for a particle with angular velocity $\omega$ is $L_{c}\sim\omega r_{c}^2$. Kepler's third law of planetary motion implies that $r_c^2 \omega^2\sim {G_{N}m}/{r_{c}}$. (Here $m$ is the mass of the central object, as above.) It therefore follows that $L_{c}\sim\sqrt{{G_Nm}/{r_{c}}}\; r_{c}$. That is $L_{c}\sim\sqrt{mr_{c}}$, as above.

Differentiating Eq.~(\ref{Lcsq}) and finding the resulting roots is not analytically feasible. We instead differentiate Eq.~(\ref{Lcsqa0}), obtaining a Taylor series for $\frac{\partial L_{c}^2}{\partial r_{c}}$ for small $a$:
\begin{eqnarray}
\frac{\partial L_{c}^2}{\partial r_{c}} &=& {mr_c(r_c-6m)\over (r_c-3m)^2} 
- {mn(5r_c-3m)a^2\over2(r_c-3m)^3}
\\
&&
-{n\{16r_c^3 +(n-2) (4 r_c^3+21mr_c^2-36m^2r_c +27m^3)  \} a^4\over 8 r_c^2 (r_c-3m)^4}
+ \O(a^6) \ .
\nonumber
\end{eqnarray}

Solving for the stationary points yields:
\begin{equation}
r_\ISCO = 6m 
\left\{ 1+ \frac{na^2}{8m^2} -{n(49n-22)a^4\over3456m^4}
+ \mathcal{O}\left(a^6\over m^6\right) \right\}\  ,
\end{equation}
and the $a=0$ Schwarzschild sanity check reproduces $r_{c} = 6m$ as required.

%-------------------------------------------------------------
\subsection{Summary}
%-------------------------------------------------------------

Denoting $r_\H$ as the location of the horizon, $r_\gamma$ as the location of the photon sphere, and $r_\ISCO$ as the location of the ISCO, we have the following summary:
\begin{itemize}
\item 
$r_{\scriptscriptstyle{H}}= 2m \times
\{1+ \frac{n a^2}{2(2m)^2}  - {n(3n-2) a^4\over 8(2m)^4} + \mathcal{O}({a^6\over m^6})\}$;
\item 
$r_\gamma= 3m \times
\left\{1 + \frac{a^2n}{(3m)^2} 
    - {n(11n-6)a^4\over 8(3m)^4} + \mathcal{O}({a^6\over m^6}) \right\}$ ;
\item 
$r_\ISCO= 6m \times
\left\{ 1+ \frac{na^2}{8m^2} -{n(49n-22)a^4\over3456m^4}+ \mathcal{O}({a^6\over m^6}) \right\} $.
\end{itemize}

%========================================
\section{Regge--Wheeler analysis}\label{Regge}
%========================================

Now considering the Regge--Wheeler equation, in view of the unified formalism developed in reference~\cite{Boonserm:2013}, (see also references~\cite{Boonserm:2018, Flachi:2012, Fernando:2012}), we may explicitly evaluate the Regge--Wheeler potentials for particles of spin $S\in\lbrace 0,1\rbrace$ in our spacetime. 
Firstly define a tortoise coordinate as follows:
\begin{equation}
\d r_{*} = {\d r\over \left(1-{a^2\over r^2}\right)^{n\over2} - {2m\over r}}.
\end{equation}
This tortoise coordinate is, for general $n$, not analytically defined. However let us make the coordinate transformation regardless; this yields the following expression for the metric:
\begin{equation}
\d s^2 = \left\{ \left(1-{a^2\over r^2}\right)^{n\over2} - \frac{2m}{r}\right\} \,
\bigg\lbrace -\d t^2+\d r_{*}^2\bigg\rbrace
+r^2\left(\d\theta^2+\sin^2\theta\; \d\phi^2\right)  .
\end{equation}
It is convenient to write this as:
\begin{equation}
\d s^2 = A(r_*)^2\bigg\lbrace -\d t^2+\d r_{*}^2\bigg\rbrace+B(r_*)^2\left(\d\theta^2+\sin^2\theta \;\d\phi^2\right)  .
\end{equation}
The Regge--Wheeler equation is~\cite{Boonserm:2013, Flachi:2012, Fernando:2012}:
\begin{equation}
\partial_{r_{*}}^{2}\hat{\phi}+\lbrace \omega^2-\mathcal{V}_S\rbrace\hat\phi = 0 ,
\end{equation}
where $\hat\phi$ is the scalar or vector field, $\mathcal{V}$ is the spin-dependent Regge--Wheeler potential for our particle, and $\omega$ is some temporal frequency component in the Fourier domain.
For a scalar field ($S=0$) examination of the d'Alembertian equation quickly yields:
\begin{equation}
\mathcal{V}_{S=0} =   \left\lbrace{A^2 \over B^2} \right\rbrace \ell(\ell+1)
+ {\partial_{r_{*}}^2 B \over B}  .
\end{equation}
For a massless vector field, ($S=1$; \emph{e.g.} photon), explicit conformal invariance in 3+1 dimensions guarantees that the Regge--Wheeler potential can depend only on the ratio $A/B$, whence normalising to known results implies:
\begin{equation}
\mathcal{V}_{S=1} =   \left\lbrace{A^2 \over B^2} \right\rbrace \ell(\ell+1).
\end{equation}
Collecting results, for $S\in\{0,1\}$ we have:
\begin{equation}
\mathcal{V}_{S\in\{0,1\}} =   \left\lbrace{A^2 \over B^2} \right\rbrace \ell(\ell+1)
+ (1-S) {\partial_{r_{*}}^2 B \over B}  .
\end{equation}
The spin 2 axial mode is somewhat messier, and (for current purposes) not of immediate interest. 

\noindent Noting that for our metric $\partial_{r_{*}} = \left\{\left(1-\frac{a^2}{r^2}\right)^{\frac{n}{2}} - \frac{2m}{r}\right\}\partial_{r}$ and $B(r)=r$  we have:
\begin{eqnarray}
\frac{\partial_{r_{*}}^2 B}{B} 
&=& \frac{\partial_{r_{*}}\left\{\left(1-\frac{a^2}{r^2}\right)^{\frac{n}{2}} - \frac{2m}{r}\right\}}{r}
\nonumber \\
&& \nonumber \\
&=&{1\over r^2} \left\{\left(1-\frac{a^2}{r^2}\right)^{\frac{n}{2}} - \frac{2m}{r}\right\}
\left\{n\left(1-\frac{a^2}{r^2}\right)^{\frac{n}{2}-1} {a^2\over r^2} + \frac{2m}{r}\right\}. \nonumber \\
\end{eqnarray}

For small $a$:
\begin{equation}
\frac{\partial_{r_{*}}^2 B}{B} = {2m(1-2m/r)\over r^3} + {n(r-3m)\over r^5}a^2 +
 {n\{5(n-2) m - 4(n-1) r\}  \over 4 r^7} a^4+ \O\left(\frac{ma^6}{r^9}\right) \ .
\end{equation}

Therefore:
\begin{eqnarray}
\mathcal{V}_{S\in\{0,1\}} &=& {1\over r^2} \left[\left(1-\frac{a^2}{r^2}\right)^{\frac{n}{2}}-\frac{2m}{r}\right]\Bigg\lbrace\ell\left(\ell+1\right)+{\left(1-S\right)}\left[n\left(1-\frac{a^2}{r^2}\right)^{\frac{n}{2}-1} \,{a^2\over r^2} + \frac{2m}{r}\right]\Bigg\rbrace. \nonumber \\
&&
\end{eqnarray}

\enlargethispage{40pt}
This has the correct behaviour as $a\to0$, reducing to the Regge--Wheeler potential for Schwarzschild:
\begin{equation}
    \lim_{a\rightarrow 0}\mathcal{V}_{S\in\lbrace 0,1\rbrace} = \frac{1}{r^2}\left[1-\frac{2m}{r}\right]\left\lbrace\ell(\ell+1)+(1-S)\frac{2m}{r}\right\rbrace \ .
\end{equation}
In the small $a$ approximation we have the asymptotic result
\begin{eqnarray}
\mathcal{V}_{S\in\lbrace 0,1\rbrace} &=& \frac{\left(1-\frac{2m}{r}\right)}{r^2}\left\lbrace\ell(\ell+1)+(1-S)\frac{2m}{r}\right\rbrace \nonumber \\
&& \nonumber \\
&& -\frac{na^{2}}{2r^4}\left\lbrace\ell(\ell+1)+2(1-S)\left[\frac{3m}{r}-1\right]\right\rbrace \nonumber \\
&& \nonumber \\
&& +\frac{na^4}{2r^6}\left\lbrace\frac{(n-2)}{4}\left[\ell(\ell+1)\right]
-\left(1-S\right)\left[2(n-1)+5\left(1-\frac{n}{2}\right)\frac{m}{r}\right]\right\rbrace + \mathcal{O}\left(\frac{a^6}{r^8}\right) \ . \nonumber \\
\end{eqnarray}
The Regge--Wheeler equation is fundamental to exploring the quasi-normal modes of the candidate spacetimes, an integral part of the ``ringdown'' phase of the LIGO calculation to detect astrophysical phenomena \emph{via} gravitational waves. Exploring the quasi-normal modes is, for now, relegated to the domain of future research.

\vspace{-10pt}
%==================================
\section{Discussion and conclusions}\label{S:conclusions}
%==================================

The original Kazakov--Solodukhin ``quantum deformed Schwarzschild spacetime''~\cite{Kazakov:1993} is 
slightly more ``regular'' than Schwarzschild spacetime, but it is not ``regular'' in the sense normally intended in the general relativity community.  While the metric components are regular, both Christoffel symbols and curvature invariants diverge at the ``centre'' of the spacetime, a 2-sphere  where $r\to a$ with finite area $A=4\pi a^2$. The ``smearing out'' of the ``centre'' to $r\to a$ is not sufficient to guarantee curvature regularity.

We have generalized the original Kazakov--Solodukhin spacetime to a two-parameter class compatible with the ideas mooted in reference~\cite{Kazakov:1993}.
Our generalized two-parameter class of ``quantum corrected'' Schwarzschild spacetimes contains exemplars which have  much better regularity properties, and we can distinguish three levels of regularity: metric regularity, Christoffel regularity, and regularity of the curvature invariants.  

\enlargethispage{40pt}
Furthermore, our generalized two-parameter class of models distorts Schwarzschild spacetime in a clear and controlled way --- so providing yet more examples of black-hole ``mimickers'' potentially of interest for observational purposes.  In this regard we have analyzed the geometry, surface gravity, stress-energy, and classical energy conditions. We have also perturbatively analyzed the locations of ISCOs and photon spheres, and set up the appropriate Regge--Wheeler formalism for spin-1 and spin-0 excitations.

Overall, the general topic of ``quantum corrected'' Schwarzschild spacetimes is certainly of significant interest, and we hope that these specific examples may serve to encourage further investigation in this field.

%========================================================
\section*{Acknowledgements}
%========================================================
TB acknowledges financial support via a MSc Masters Scholarship provided by Victoria University of Wellington. TB is also indirectly supported by the Marsden fund, administered by the Royal Society of New Zealand.\\
AS acknowledges financial support via a PhD Doctoral Scholarship provided by Victoria University of Wellington. AS is also indirectly supported by the Marsden fund, administered by the Royal Society of New Zealand.\\
MV was directly supported by the Marsden Fund, via a grant administered by the Royal Society of New Zealand.

%\clearpage
%========================================================
%========================================================

%========================================================
\end{document}